\documentclass[aps,prd,nofootinbib,amsmath,amssymb,showpacs,superscriptaddress,showkeys,twocolumn,10pt]{revtex4}
\usepackage{txfonts}
\usepackage{graphicx}
\usepackage{dcolumn}
\usepackage{bm}
\usepackage{amssymb}
\usepackage{latexsym}
\usepackage{booktabs}
\usepackage{amsmath}
\usepackage{multirow}
\usepackage[colorlinks=true, linkcolor=red, citecolor=blue]{hyperref}

\newcommand{\be}{\begin{equation}}
\newcommand{\ee}{\end{equation}}
\newcommand{\bq}{\begin{eqnarray}}
\newcommand{\eq}{\end{eqnarray}}

\begin{document}

\title{Revisiting the interacting model of new agegraphic dark energy}

\author{Jing-Fei Zhang}
\affiliation{College of Sciences, Northeastern University, Shenyang
110004, China}
\author{Li-Ang Zhao}
\affiliation{College of Sciences, Northeastern University, Shenyang
110004, China}
\author{Xin Zhang\footnote{Corresponding author}}
\email{zhangxin@mail.neu.edu.cn} \affiliation{College of Sciences,
Northeastern University, Shenyang 110004, China} \affiliation{Center
for High Energy Physics, Peking University, Beijing 100080, China}

\begin{abstract}
In this paper, a new version of the interacting model of new
agegraphic dark energy (INADE) is proposed and analyzed in detail.
The interaction between dark energy and dark matter is reconsidered.
The interaction term $Q=bH_0\rho_{\rm de}^\alpha\rho_{\rm
dm}^{1-\alpha}$ is adopted, which abandons the Hubble expansion rate
$H$ and involves both $\rho_{\rm de}$ and $\rho_{\rm dm}$. Moreover,
the new initial condition for the agegraphic dark energy is used,
which solves the problem of accommodating baryon matter and
radiation in the model. The solution of the model can be given using
an iterative algorithm. A concrete example for the calculation of
the model is given. Furthermore, the model is constrained by using
the combined Planck data (Planck+BAO+SNIa+$H_0$) and the combined
WMAP-9 data (WMAP+BAO+SNIa+$H_0$). Three typical cases are
considered: (A) $Q=bH_0\rho_{\rm de}$, (B) $Q=bH_0\sqrt{\rho_{\rm
de}\rho_{\rm dm}}$, and (C) $Q=bH_0\rho_{\rm dm}$, which correspond
to $\alpha=1$, 1/2, and 0, respectively. The departures of the
models from the $\Lambda$CDM model are measured by the $\Delta$BIC
and $\Delta$AIC values. It is shown that the INADE model is better
than the NADE model in the fit, and the INADE(A) model is the best
in fitting data among the three cases.
\end{abstract}


 \maketitle

Cosmological observations continue to indicate that the universe is
currently experiencing an accelerated
expansion~\cite{sn98,wmap9,planck}. This cosmic acceleration is
commonly believed to be caused by ``dark energy,'' something
producing gravitational repulsion. However, the nature of dark
energy, hitherto, is still unknown.

The simplest candidate for dark energy is Einstein's cosmological
constant, $\Lambda$, which is equivalent to the vacuum energy
density in the universe and produces negative pressure with $w=-1$
(here, $w$ is the equation of state parameter, defined by
$w=p/\rho$). However, the cosmological constant is theoretically
challenged: its observationally required value is $10^{120}$ times
smaller than its theoretical expectation. So, cosmologists are in
need of new theoretical insights. Alternatively, dark energy might
be due to some unknown scalar field, usually dubbed quintessence,
that could supply the requisite negative pressure to accelerate the
cosmic expansion. Scalar field is, nevertheless, only one option.
Many dynamical dark energy models from various theoretical
perspectives have been proposed; for reviews, see, e.g.,
Refs.~\cite{dereview}. In particular, there is an attractive idea
that links the vacuum energy density with the holographic principle
of quantum gravity. This class of models is called ``holographic
dark energy,'' in which the UV problem of dark energy is converted
to an IR problem and the dark energy density can be expressed as
$\rho_{\rm de}\propto L^{-2}$ where $L$ is the IR length-scale
cutoff of the theory. As a consequence of the effective quantum
field theory, the vacuum energy density in this theory is not a
constant, but dynamically evolutionary.

The original version of the holographic dark energy chooses the
event horizon size of the universe as the IR cutoff~\cite{Li:2004};
for extensive studies of this model, see, e.g.,
Refs.~\cite{holoext,holofit}. Subsequently, other versions were
proposed; for example, the so called ``agegraphic dark energy''
model (here, we refer to the new agegraphic dark energy model,
abbreviated as NADE) chooses the conformal time (age) of the
universe as the IR cutoff~\cite{Wei:2007nade}. Thus, in this model,
the dark energy density is of the form,
\begin{equation}
\rho_{\rm de}=3n^2M_{\rm Pl}^2\eta^{-2},
\end{equation}
where $n$ is a numerical parameter, $M_{\rm Pl}$ is the reduced
Planck mass, and $\eta$ is the conformal age of the universe,
\begin{equation}
\eta\equiv\int_0^t\frac{dt}{a}=\int_0^a\frac{da}{Ha^2}.
\end{equation}

Actually, this model can also be derived from the uncertainty
relation of quantum mechanics together with the gravitational effect
in general relativity~\cite{Cai:2007ade}. The most attractive merit
of the NADE model is that it has the same number of parameters as
the $\Lambda$CDM (the cosmological constant plus cold dark matter)
model, less than other dynamical dark energy models. The model has
also been proven to fit the data well~\cite{nadefit}. See also,
e.g., Refs.~\cite{NADEext,NADEini} for various studies of the NADE
model.

It seems necessary to consider the important possibility that there
is some direct interaction between dark energy
and dark matter. Though there is no convincing observational
evidence for this coupling, such a hypothesis has inspired considerable
theoretical interests. A large number of interacting dark energy
models have been investigated. The interacting model of new
agegraphic dark energy (INADE) was proposed and studied in detail in
Refs.~\cite{INADEold1,INADEold2}. If dark energy interacts with cold
dark matter, the continuity equations for them are written as
\begin{eqnarray}
  \dot \rho_{\rm de}+3H(1+w)\rho_{\rm de} &=& -Q,\label{continuity1}\\
  \dot \rho_{\rm dm}+3H     \rho_{\rm dm} &=&  Q,\label{continuity2}
\end{eqnarray}
where $w$ is the equation of state parameter of dark energy, and $Q$
phenomenologically describes the interaction between dark energy and
dark matter. Since we have no a fundamental theory to determine the
form of $Q$, its form needs to be assumed phenomenologically.
The most common choice is $Q\propto H\rho$, where $\rho$ denotes the
density of dark energy or dark matter (or the sum of the two). Such
a scenario is mathematically simple, but it is difficult to see how this form can emerge from a physical
description of dark sector interaction. It is expected that the interaction is determined by
the local properties of the dark sectors, i.e., $\rho_{\rm de}$ and
$\rho_{\rm dm}$, but it is hard to understand why the
interaction term must be proportional to the Hubble expansion rate $H$. 
A more natural hypothesis is
that the Hubble parameter is abandoned and thus $Q$ is only
proportional to the dark sector density, namely, $Q\propto \rho_{\rm
de}$ or $Q\propto \rho_{\rm dm}$. Such a scenario has also been
studied widely; see, e.g., Refs.~\cite{intde}.

In Ref.~\cite{INADEold2} the INADE model was investigated in detail,
but there are some issues that should be re-scrutinized under the
current situation: (i) In Ref.~\cite{INADEold2} the interaction term
is assumed to be of the form $Q\propto H\rho$; but now, it seems
necessary to adopt the more natural form in which the Hubble
parameter $H$ is abandoned. (ii) In Ref.~\cite{INADEold2} the model
can only accommodate two components, dark energy and dark matter,
but cannot involve baryon matter, radiation and spatial curvature.
This is caused by the old initial condition used. In
Ref.~\cite{NADEini}, however, a new initial condition (with a
numerical algorithm) of the model was proposed, which can solve this
problem. (iii) Recently, the Planck Collaboration publicly released
the cosmic microwave background (CMB) temperature and lensing
data~\cite{planck}, so it is also necessary to constrain the model
by using the new data. Thus, under the current situation, we revisit
the INADE model in this paper.

In this work, we will consider a more general form for the
interaction term. Following Ref.~\cite{intholo} in which the
interacting model of holographic dark energy was discussed in
detail, we will take the form
\begin{equation}
Q \propto \rho_{\rm de}^{\alpha}\rho_{\rm dm}^{\beta},\label{int}
\end{equation}
which includes the forms $Q\propto\rho_{\rm de}$ and
$Q\propto\rho_{\rm dm}$ as special cases, describing the decay
process of dark energy or dark matter. Moreover, this form can also
describe the more complicated cases of interaction, such as
scattering, in which one may expect the existence of both $\rho_{\rm
de}$ and $\rho_{\rm dm}$. Therefore, Eq.~(\ref{int}) is,
undoubtedly, a more natural and physically plausible form in
describing the interaction between dark energy and dark matter. In
the following, for simplicity, we confine our discussion in the
class with the condition $\alpha+\beta=1$, so the interaction term
can be explicitly expressed as
\begin{equation}
Q=bH_0\rho_{\rm de}^\alpha\rho_{\rm dm}^{1-\alpha},\label{intQ}
\end{equation}
where $b$ is the coupling constant; when $b>0$, the energy flow is
from dark energy to dark matter, and when $b<0$ the energy flow is
from dark matter to dark energy.

In what follows we will discuss the INADE model with the interaction
form (\ref{intQ}). We will use the new initial condition proposed in
Ref.~\cite{NADEini}. After the numerical solution of the model is
given, we will further constrain the parameter space of the model by
using the latest observational data, including the CMB data, the
baryon acoustic oscillation (BAO) data, the type Ia supernova (SNIa)
data, and the Hubble constant data. In particular, for the CMB data,
we will use both the Planck data~\cite{planck} and the 9-year
Wilkinson Microwave Anisotropy Probe (WMAP-9) data~\cite{wmap9},
respectively, for a comparison.

Defining $f_{\rm de}\equiv \rho_{\rm de}/\rho_{\rm de0}$ and $f_{\rm
dm}\equiv \rho_{\rm dm}/\rho_{\rm dm0}$ (the subscript ``0'' in this
paper denotes the present value of the corresponding quantity), we
can rewrite Eqs.~(\ref{continuity1}) and (\ref{continuity2}) into
the following forms,
\begin{eqnarray}
  \frac{df_{\rm de}(x)}{dx}+3(1+w_{\rm de, eff})f_{\rm de}(x)&=&0, \label{maineq1}\\
  \frac{df_{\rm dm}(x)}{dx}+3(1+w_{\rm dm, eff})f_{\rm dm}(x)&=&0, \label{maineq2}
\end{eqnarray}
where $x=\ln a$, and
\begin{eqnarray}
w_{\rm de, eff}(x) &=& -1+\frac{2}{3 n e^x E(x)} \sqrt{\Omega_{\rm de0} f_{\rm de}(x)}, \\
w_{\rm dm, eff}(x) &=&-{b r^\alpha\over3E(x)}f_{\rm de}(x)^\alpha
f_{\rm dm}(x)^{-\alpha},
\end{eqnarray}
with $r\equiv\rho_{\rm de0}/\rho_{\rm dm0}$ and $E(x)\equiv
H(x)/H_0$. Note that in this paper we consider a flat universe, so
the Friedmann equation $3M_{\rm Pl}^2H^2=\rho_{\rm de}+\rho_{\rm
dm}+\rho_{\rm b}+\rho_{\rm r}$ can be recast as
\begin{equation}
E(x)=\sqrt{\Omega_{\rm de0} f_{\rm de}(x)+\Omega_{\rm dm0} f_{\rm
dm}(x)+\Omega_{\rm b0} e^{-3x}+\Omega_{\rm r0} e^{-4x}}.
\end{equation}
The solutions to the system of differential equations
(\ref{maineq1}) and (\ref{maineq2}), $f_{\rm de}(x)$ and $f_{\rm
dm}(x)$, completely describe the cosmological evolution of the INADE
model. Hence, next, the main task is to find out the solutions to
Eqs. (\ref{maineq1}) and (\ref{maineq2}). In order to solve the
differential equations, we first need to give the initial conditions
for them.

As shown in Ref.~\cite{Wei:2007nade}, the parameters $\Omega_{\rm
m0}$ and $n$ are not independent of each other. Once $n$ is given,
$\Omega_{\rm m0}$ can be derived, and vice versa. If one takes both
$\Omega_{\rm m0}$ and $n$ as free parameters, the NADE model will
become problematic; see Fig.~2~(a) in Ref.~\cite{INADEold2} and the
corresponding discussions. So the initial conditions in this model
should be taken at the early times; usually, as a convention, the
initial conditions are taken at $z_{\rm ini}=2000$, in the
matter-dominated epoch. Since in the matter-dominated epoch the
contribution of dark energy to the cosmological evolution is
negligible, it is expected that the impact of the interaction on
dark energy in the early times is also ignorable. So, it is suitable
to follow Ref.~\cite{NADEini} to take the initial condition for the
agegraphic dark energy in this model as
\begin{equation}
f_{\rm de}(x_{\rm ini})= \frac{n^2 \Omega_{\rm m0}^2}{4 \Omega_{\rm
de0}} (\sqrt{\Omega_{\rm m0}e^{x_{\rm ini}}+\Omega_{\rm
r0}}-\sqrt{\Omega_{\rm r0}})^{-2}, \label{cond2}
\end{equation}
where $\Omega_{\rm m0}=\Omega_{\rm dm0}+\Omega_{\rm b0}$. In fact,
the impact of interaction on dark matter in the early times is also
fairly small, so the deviation from the scaling law $a^{-3}$ for
dark matter is tiny. Nevertheless, we still use a small quantity
$\delta$ to parameterize this tiny deviation; so the initial
condition for dark matter is taken as
\begin{equation}
f_{\rm dm}(x_{\rm ini})=e^{(-3+\delta)x_{\rm ini}}.\label{cond1}
\end{equation}
We shall show that $\delta$ is also a derived parameter, and its
value can be determined using an iteration calculation. Throughout
the calculation, we fix $\Omega_{\rm r0}=2.469\times10^{-5}
h^{-2}(1+N_{\rm eff})$, where $N_{\rm eff}=3.046$ is the standard
value of the effective number of the neutrino species, and $h$ is
the Hubble constant $H_0$ in units of 100 km s$^{-1}$ Mpc$^{-1}$.

\begin{figure*}[htbp]
\centering
\begin{center}
$\begin{array}{c@{\hspace{0.6in}}c} \multicolumn{1}{l}{\mbox{}} &
\multicolumn{1}{l}{\mbox{}} \\
\includegraphics[scale=0.35]{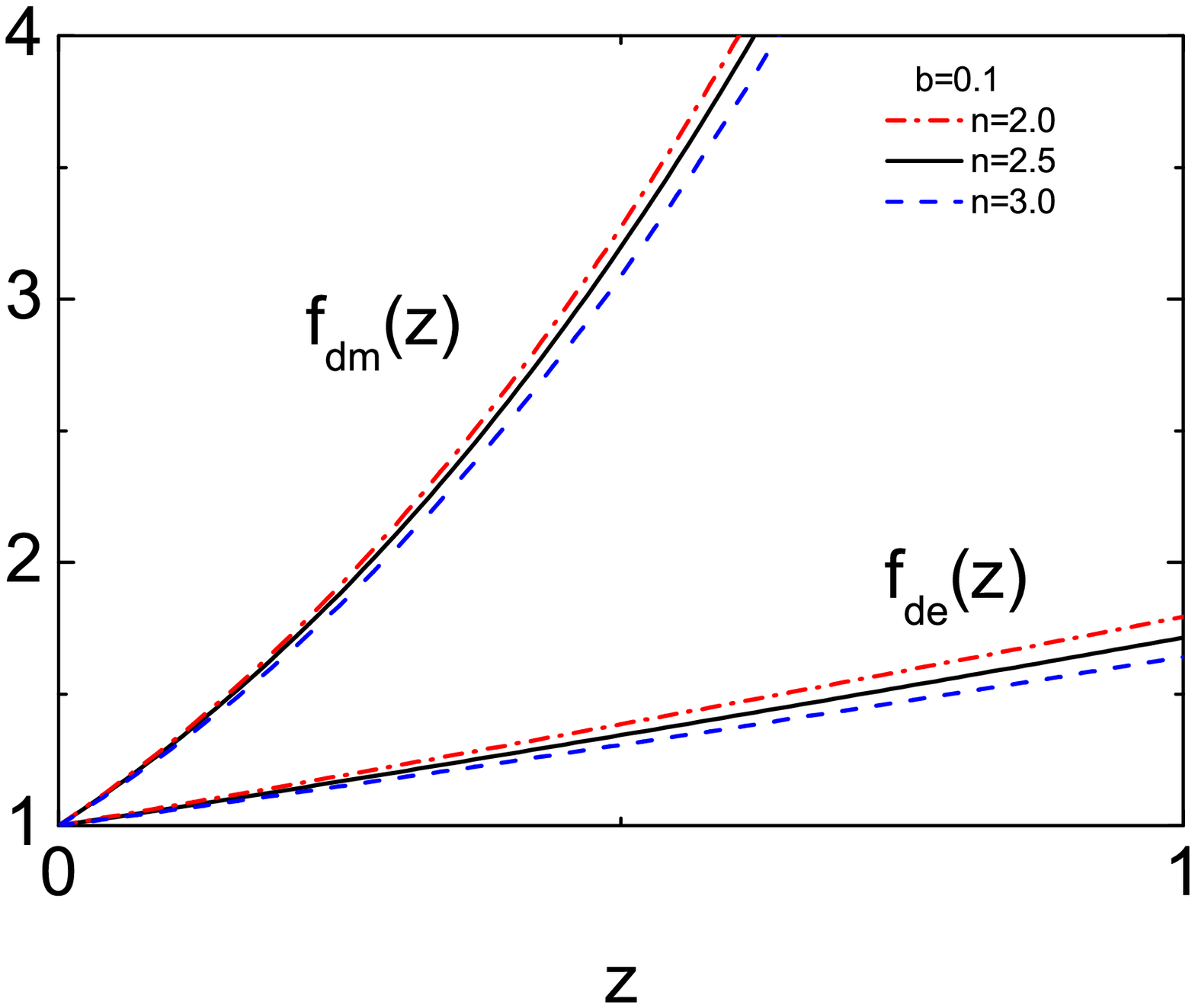} &\includegraphics[scale=0.35]{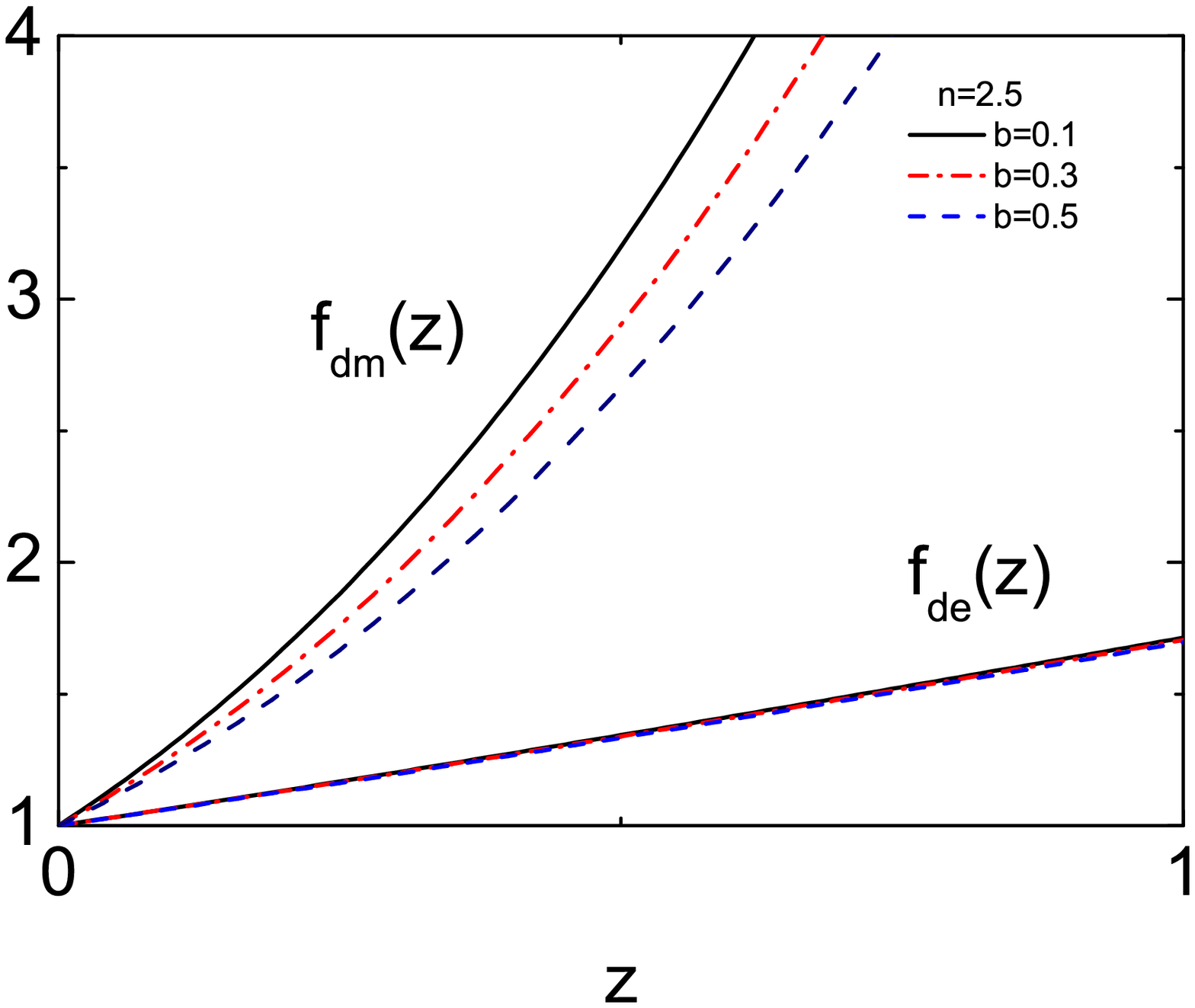} \\
\end{array}$
\end{center}
\caption[]{\small \label{fig:example}The solutions of
Eqs.~(\ref{maineq1}) and (\ref{maineq2}), $f_{\rm de}(z)$ and
$f_{\rm dm}(z)$, in the INADE model with $Q=bH_0\rho_{\rm de}$. In
this case, we fix $h=0.7$ and $\Omega_{\rm b0}=0.05$. In the left
panel, we fix $b=0.1$ and vary $n$; in the right panel, we fix
$n=2.5$ and vary $b$.}
\end{figure*}

In this model, the free parameters are: $n$, $b$, $\Omega_{\rm b0}$,
and $h$. The independent parameters $\delta$ and $\Omega_{\rm m0}$
(or $\Omega_{\rm dm0}$) can be derived using an iterative algorithm.
In our calculation, we employ the Newton iteration method. The
conditions of convergence we set are:
$\frac{\delta^{(l+1)}-\delta^{(l)}}{\delta^{(l)}}<10^{-5}$ and
$\frac{\Omega_{\rm m0}^{(l+1)}-\Omega_{\rm m0}^{(l)}}{\Omega_{\rm
m0}^{(l)}}<10^{-5}$, where the index $l$ denotes the iteration times
in the numerical calculation.

To show the solution of the model, we give a concrete example. In
this example, we take $\alpha=1$ in Eq.~(\ref{intQ}), i.e.,
$Q=bH_0\rho_{\rm de}$. Furthermore, we fix $h=0.7$ and $\Omega_{\rm
b0}=0.05$, in order to explicitly show the impacts of the parameters
$n$ and $b$ on the model. The solutions $f_{\rm de}(z)$ and $f_{\rm
dm}(z)$ are shown in Fig.~\ref{fig:example}. In the left panel, we
fix $b=0.1$, and take $n=2.0$, 2.5 and 30., respectively; one can
see that the parameter $n$ impacts both evolutions of dark
matter and dark energy evidently. In the right panel, we fix
$n=2.5$, and take $b=0.1$, 0.3 and 0.5, respectively; one can see
that the coupling strength $b$ impacts on the evolution of dark
matter more evidently than the evolution of dark energy. The
parameters $\Omega_{\rm m0}$ and $\delta$ can be derived through the
iterative calculation. For instance, for the case $n=2.5$ and
$b=0.1$, we obtain $\Omega_{\rm m0}=0.340$ and $\delta=0.010$.
Actually, the calculations for all the cases show that $\delta$ is
around ${\cal O}(10^{-2})$, verifying the previous statement for the
initial condition of dark matter that the early deviation from the
scaling law $a^{-3}$ for dark matter is indeed tiny. Obviously, once
the solutions to Eqs.~(\ref{maineq1}) and (\ref{maineq2}), $f_{\rm
de}(z)$ and $f_{\rm dm}(z)$, are given, the other quantities of
interest, such as $\Omega_{\rm de}(z)$, $\Omega_{\rm dm}(z)$,
$w(z)$, $H(z)$, $Q(z)$, and so on, can be directly calculated.
Thus, so far, we have proposed the revised version of the INADE
model and given the solution of this model. Since the issues such as
the alleviation of the cosmic coincidence problem are the common
characteristics of interacting dark energy models, we do not discuss
this class of issues in this paper. Next, we will test this model
with the latest observational data and explore the parameter space
of the model in the fit to data.

\begin{table*}[htbp]
\caption{\label{table1}The fit results of the $\Lambda$CDM, NADE,
and INADE models. For the INADE model, cases with (A)
$Q=bH_0\rho_{\rm de}$, (B) $Q=bH_0\sqrt{\rho_{\rm de}\rho_{\rm
dm}}$, and (C) $Q=bH_0\rho_{\rm dm}$ are considered. Since the
numbers of parameters are different for different models, the
information criteria BIC and AIC are used in the model comparison.
Here, $k$ denotes the number of parameters in the models. The order
of the three cases of the INADE model is arranged according to the
fit results. In the fit, we use the data combination
CMB+BAO+SNIa+$H_0$, where for CMB we use the Planck data and WMAP-9
data, respectively, for a comparison. The best-fit values with
1--2$\sigma$ errors for the parameters in the models are presented.}
\begin{tabular}{c|c|c|c|c|c|c|c|c|c|c}
\hline\hline
Data & Model & $\Omega_{\rm m0}$ & $\Omega_{\rm b0}$ & $n$ & $h$ & $b$ & $k$ & $\chi_{\rm min}^2$ & $\Delta$BIC & $\Delta$AIC\\
\hline \multirow{5}{*}{Planck}
  & $\Lambda$CDM    & $0.298^{+0.014+0.024}_{-0.013-0.022}$ & $0.047^{+0.001+0.002}_{-0.001-0.002}$ &
  -
  & $0.688^{+0.011+0.018}_{-0.011-0.018}$ & -
  & $3$ & $554.138$ &   $0$    & $0$     \\

  &     NADE        & - & $0.058^{+0.001+0.002}_{-0.001-0.002}$ & $2.471^{+0.069+0.115}_{-0.070-0.115}$
  & $0.633^{+0.009+0.015}_{-0.009-0.014}$ & -
  & $3$ & $589.309$ & $35.171$ & $35.171$ \\

  &    INADE(A)       & - & $0.052^{+0.002+0.004}_{-0.002-0.003}$ & $2.609^{+0.083+0.083}_{-0.139-0.136}$
  & $0.658^{+0.012+0.020}_{-0.012-0.020}$ & $0.088^{+0.025+0.042}_{-0.026-0.045}$
  & $4$ & $568.312$ & $20.552$ & $16.174$ \\

  &    INADE(B)       & - & $0.052^{+0.002+0.004}_{-0.002-0.003}$ & $2.602^{+0.083+0.137}_{-0.083-0.138}$
  & $0.657^{+0.012+0.020}_{-0.012-0.020}$ & $0.096^{+0.029+0.047}_{-0.030-0.051}$
  & $4$ & $569.133$ & $21.373$ & $16.995$ \\

  &    INADE(C)       & - & $0.052^{+0.002+0.003}_{-0.002-0.003}$  & $2.592^{+0.081+0.136}_{-0.082-0.133}$
  & $0.654^{+0.012+0.020}_{-0.012-0.019}$ & $0.086^{+0.028+0.047}_{-0.029-0.048}$
  & $4$ & $570.387$ & $22.627$ & $18.249$ \\
\hline \multirow{5}{*}{WMAP-9}
  & $\Lambda$CDM    & $0.295^{+0.016+0.026}_{-0.015-0.024}$ & $0.048^{+0.002+0.003}_{-0.001-0.002}$ &
  -
  & $0.690^{+0.013+0.022}_{-0.013-0.021}$ & -
  & $3$ & $554.128$ & $0$       & $0$     \\

  &     NADE        & - & $0.056^{+0.002+0.003}_{-0.002-0.003}$ & $2.580^{+0.082+0.135}_{-0.082-0.135}$
  & $0.647^{+0.012+0.019}_{-0.011-0.019}$ & -
  & $3$ & $576.168$ & $22.040$ & $22.040$ \\

  &    INADE(A)       & - & $0.052^{+0.002+0.004}_{-0.002-0.004}$ & $2.621^{+0.086+0.143}_{-0.085-0.140}$
  & $0.659^{+0.013+0.022}_{-0.013-0.022}$ & $0.081^{+0.042+0.069}_{-0.041-0.070}$
  & $4$ & $567.993$ & $20.243$ & $15.865$ \\

  &    INADE(B)       & - & $0.053^{+0.002+0.004}_{-0.002-0.004}$ & $2.616^{+0.084+0.141}_{-0.084-0.140}$
  & $0.658^{+0.013+0.022}_{-0.013-0.021}$ & $0.087^{+0.045+0.075}_{-0.046-0.078}$
  & $4$ & $568.737$ & $20.987$ & $16.609$ \\

  &   INADE(C)       & - & $0.053^{+0.002+0.004}_{-0.002-0.004}$ & $2.608^{+0.084+0.140}_{-0.085-0.140}$
  & $0.655^{+0.013+0.021}_{-0.012-0.020}$ & $0.075^{+0.043+0.070}_{-0.044-0.074}$
  & $4$ & $569.832$ & $22.082$ & $17.704$ \\
\hline
\end{tabular}
\end{table*}

\begin{figure*}[htbp]
\centering
\begin{center}
$\begin{array}{c@{\hspace{0.6in}}c} \multicolumn{1}{l}{\mbox{}} &
\multicolumn{1}{l}{\mbox{}} \\
\includegraphics[scale=0.35]{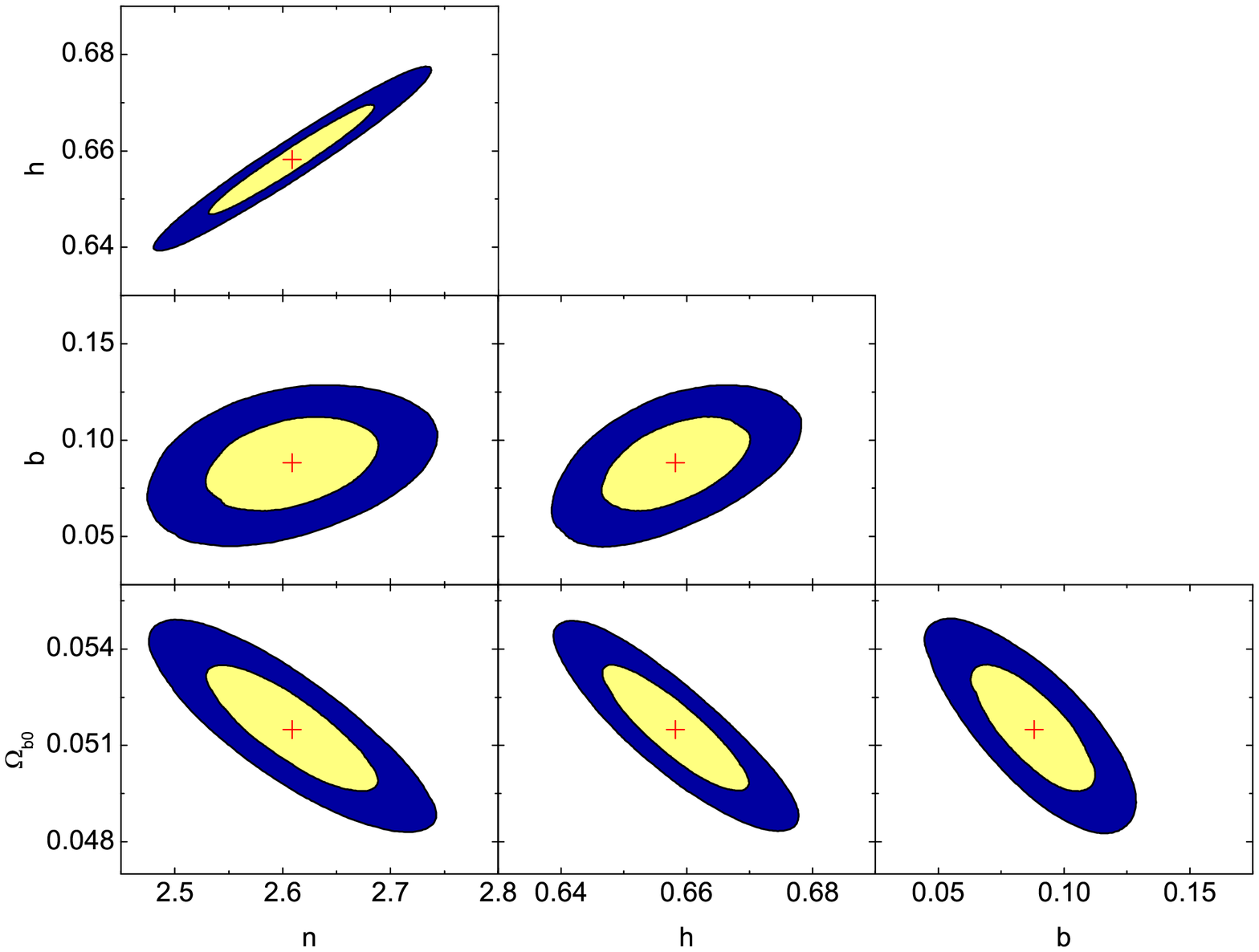} &\includegraphics[scale=0.35]{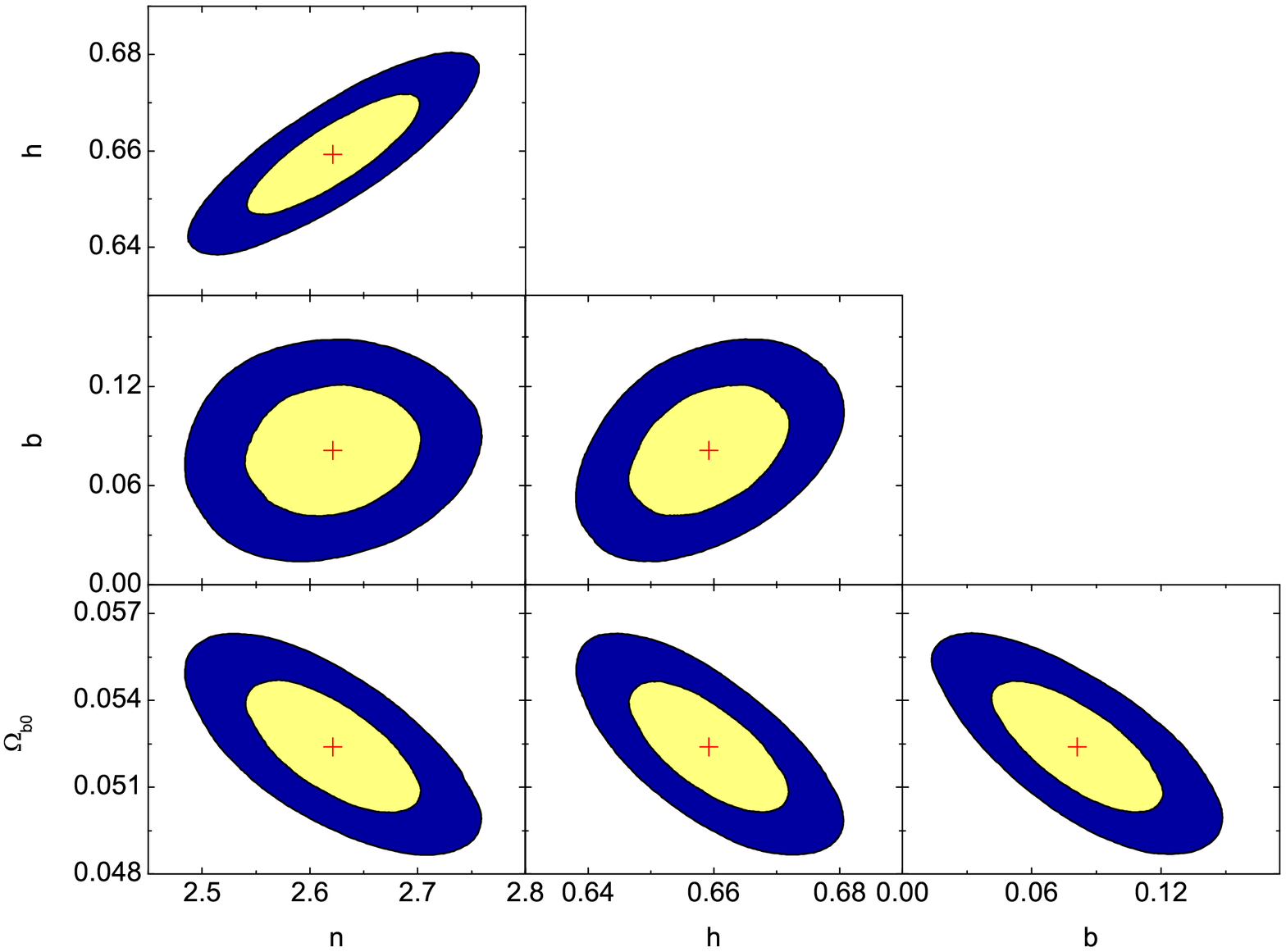} \\
\end{array}$
\end{center}
\caption[]{\small \label{fig:modelA}The two-dimensional marginalized
constraints ($68\%$ and $95\%$ CLs) on the INADE(A) model with
$Q=bH_0\rho_{\rm de}$, from the CMB+BAO+SNIa+$H_0$ data. The CMB
data used in the two panels are different: in the left panel, the
Planck data are used; in the right panel, the WMAP-9 data are used.}
\end{figure*}

In the fit, we only focus on several typical cases of the model. The
cases we consider include: (A) $Q=bH_0\rho_{\rm de}$, (B)
$Q=bH_0\sqrt{\rho_{\rm de}\rho_{\rm dm}}$, and (C) $Q=bH_0\rho_{\rm
dm}$, which correspond to $\alpha=1$, 1/2, and 0, respectively.

The data we use include the CMB data, BAO data, SNIa data, and
Hubble constant data. For CMB, we use both the Planck
data~\cite{planck} and the WMAP-9 data~\cite{wmap9}, for a
comparison. In this work, we do not consider the cosmological
perturbations in the calculation. This avoids the extra assumptions
on the sound speed of dark energy perturbation, the momentum
transport between dark energy and dark matter, and so forth. In
fact, the result will not be affected evidently when the
cosmological perturbations are involved. Since the perturbations are
ignored, we can use the CMB distance prior data in the fit. The
results of the CMB distance priors ($l_{\rm A}$, $R$, $\omega_{\rm
b}$) for Planck and WMAP-9 have been given in
Ref.~\cite{Wang:2013mha}, so we will use these data in our fit
analysis. For BAO, we use the SDSS-DR7, SDSS-DR9, 6dFGS, and
WiggleZ data; the prescription of the use of these data has been
given in Ref.~\cite{wmap9}. For SNIa, we use the Union2.1
data~\cite{union2.1}. For the Hubble constant measurement, we use
the HST result~\cite{Riess:2011HST}, $H_0=73.8\pm 2.4$ km s$^{-1}$
Mpc$^{-1}$. The Markov Chain Monte Carlo (MCMC) method is employed
in our data fit analysis.

The fit results are presented in Table~\ref{table1}. Since the
$\Lambda$CDM model fits the data very well, in this work we also
make a comparison with the $\Lambda$CDM model. Actually, the
$\Lambda$CDM model has been adopted as a fiducial model in dark
energy cosmology. In addition, a comparison with the NADE model
without interaction is also made. The numbers of parameters in the
$\Lambda$CDM model and the NADE model are equal, but they are less
than that in the INADE model. We denote the number of parameters in a
model as $k$. For the $\Lambda$CDM model and the NADE model, $k=3$;
For the INADE model (with $\alpha$ fixed), $k=4$. In order to fairly
compare the models with different numbers of parameters, we employ
the information criteria (IC), such as the Bayesian information
criterion (BIC) and the Akaike information criterion (AIC), as the
assessment tools. They are defined as ${\rm BIC}=\chi^2_{\rm min}+k
\ln N$ and ${\rm AIC}=\chi^2_{\rm min}+2k$, where $N$ is the number
of the data used in the fit. Statistically, a model with few
parameters and with a better fit to the data has lower IC values.
Thus, the models can be ranked according to their IC values. 
Note that the parameter $h$ is included in the number of degrees of 
freedom and in $k$ as a parameter in each model, since it appears in 
the fits to the data of CMB, BAO, and $H_0$, and cannot be marginalized 
in the fits.
Also, in this
work, the number of data $N$ is fixed, so the BIC and the AIC
produce the same order of the models. The $\Lambda$CDM model is
proven to be the best in fitting data (i.e., with lowest values of
BIC and AIC; see also Refs.~\cite{demodels,Wei:2010BIC}), so in
Table~\ref{table1} the $\Delta$BIC and $\Delta$AIC values are
measured with respect to the $\Lambda$CDM model. In this table we
list the free parameters for the models, and present their best fit
values with 1--2$\sigma$ errors. For both the combined Planck data
and the combined WMAP-9 data, the $\Lambda$CDM model performs best,
and the NADE model performs worst. Among the three cases of the
INADE model, the case (A) is the best in fitting data, though the
difference between them is little. The parameter $\Omega_{\rm m0}$
can be derived in the INADE model. For example, for the combined
Planck data, the best-fit values of $\Omega_{\rm m0}$ are 0.317,
0.318 and 0.319, for the cases (A), (B) and (C), respectively. Also,
we find that the coupling constant $b$ in the INADE models is always
positive, indicating that the energy transport is from dark energy
to dark matter. This is helpful in alleviating the cosmic
coincidence problem.

The parameter spaces of the INADE models are also explored. Since
the INADE(A) model (with $Q=bH_0\rho_{\rm de}$) is the best in
fitting data among the three interacting models, we only show the
parameter space of this model in this paper. In fact, the three
cases are similar, so the two-dimensional contours in the parameter
planes for the INADE(B) and INADE(C) models are not reported; for a
full report on these results, see Ref.~\cite{zhao2013}. In
Fig.~\ref{fig:modelA} we show the two-dimensional marginalized
contours ($68\%$ and $95\%$ CLs) in the parameter planes for the
INADE(A) model. The left panel is the case for the combined Planck
data, and the right panel is for the combined WMAP-9 data. The
degeneracies between the parameters can be explicitly seen from this
figure. We can see that, for the Planck case, the parameter space is
more tight, but the degeneracies are more evident, especially for
the $n$--$h$ plane. Since $\Omega_{\rm m0}$ is derived from $n$,
this implies that $h$ degenerates strongly with $\Omega_{\rm m0}$,
which is consistent with the case of the $\Lambda$CDM
model~\cite{planck,Li:2013hdePlanck}.

In summary, a revised version of the INADE model is proposed and
analyzed in this paper. In this version, the interaction between
dark energy and dark matter is reconsidered. The interaction term
$Q=bH_0\rho_{\rm de}^\alpha\rho_{\rm dm}^{1-\alpha}$ is adopted,
which abandons the Hubble expansion rate $H$ and involves both
$\rho_{\rm de}$ and $\rho_{\rm dm}$. Moreover, in this version the
new initial condition for the agegraphic dark energy is used, which
solves the problem of accommodating baryon matter and radiation in
the model. The solution of the model can be given using an iterative
algorithm. We give a concrete example for the calculation of the
model. Furthermore, we constrain the model by using the combined
Planck data (Planck+BAO+SNIa+$H_0$) and the combined WMAP-9 data
(WMAP+BAO+SNIa+$H_0$). We focus on the three typical cases: (A)
$Q=bH_0\rho_{\rm de}$, (B) $Q=bH_0\sqrt{\rho_{\rm de}\rho_{\rm
dm}}$, and (C) $Q=bH_0\rho_{\rm dm}$, which correspond to
$\alpha=1$, 1/2, and 0, respectively. The departures of the models
from the $\Lambda$CDM model are measured by the $\Delta$BIC and
$\Delta$AIC values. We show that the INADE model is better than the
NADE model in the fit, and the INADE(A) model is the best in fitting
data among the three cases. As an example, we show the
two-dimensional marginalized contours ($68\%$ and $95\%$ CLs) in the
parameter planes for the INADE(A) model. It is indicated that, for
the Planck case, the parameter space is more tight, but the
degeneracies between the parameters are more evident, especially for
the $n$--$h$ plane. Owing to the fact that $\Omega_{\rm m0}$ is
derived from $n$, this implies that $h$ degenerates strongly with
$\Omega_{\rm m0}$, which is consistent with the case of the
$\Lambda$CDM model.

\begin{acknowledgments}
This work was supported by the National Natural Science Foundation of
China (Grants No.~10975032 and No.~11175042), the National Ministry
of Education of China (Grants No.~NCET-09-0276, No.~N110405011 and
No.~N120505003), and the Provincial Department of Education of
Liaoning (Grant No.~L2012087).
\end{acknowledgments}


\begin{thebibliography}{99}

\bibitem{sn98}
  Riess A G, Filippenko A V, Challis P, et al. [Supernova Search Team Collaboration]
Observational evidence from supernovae for an accelerating
universe and a cosmological constant. Astron J, 1998, 116: 1009-1038;
Perlmutter S, Aldering G, Goldhaber G, et al.
[Supernova Cosmology Project Collaboration] Measurements of $\Omega$ and
$\Lambda$ from 42 high-redshift supernovae. Astrophys J, 1999, 517: 565-586.
\bibitem{wmap9}
  Hinshaw G, Larson D, Komatsu E, et al. [WMAP Collaboration]
  Nine-year Wilkinson Microwave Anisotropy Probe (WMAP) observations: Cosmological parameter results.
  arXiv:1212.5226.

\bibitem{planck}
   Ade P A R, Aghanim N, Armitage-Caplan C, et al.  [Planck Collaboration]
  Planck 2013 results. XVI. Cosmological parameters.
  arXiv:1303.5076.

\bibitem{dereview}
Weinberg S. The cosmological constant problem.
  Rev Mod Phys, 1989, 61: 1-23;
Sahni V, Starobinsky A A.
  The case for a positive cosmological Lambda-term.
  Int J Mod Phys D, 2000, 9: 373-444;
Padmanabhan T.
  Cosmological constant: The weight of the vacuum.
  Phys Rept, 2003, 380: 235-320;
Peebles P J E, Ratra B.
  The cosmological constant and dark energy.
  Rev Mod Phys, 2003, 75: 559-606;
Copeland E J, Sami M, Tsujikawa S.
 Dynamics of dark energy.
  Int J Mod Phys D, 2006, 15: 1753-1935;
  Frieman J, Turner M, Huterer D.
 Dark energy and the accelerating universe.
  Ann Rev Astron Astrophys, 2008, 46: 385-432;
  Li M, Li X D, Wang S, Wang Y.
  Dark energy.
Commun Theor Phys, 2011, 56: 525-604.


\bibitem{Li:2004}
  Li M.
 A model of holographic dark energy.
  Phys Lett B, 2004, 603: 1-5.

\bibitem{holoext}
  Huang Q G, Li M.
  Anthropic principle favors the holographic dark energy.
  J Cosmol Astrpart Phys, 2005, 03: 001;
 Zhang X.
  Statefinder diagnostic for holographic dark energy model.
  Int J Mod Phys D, 2005, 14: 1597-1606;
Chen B, Li M, Wang Y.
 Inflation with holographic dark energy.
  Nucl Phys B, 2007, 774: 256-267;
Zhang X. Reconstructing
holographic quintessence. Phys Lett B, 2007, 648: 1-7;
Zhang X. Dynamical vacuum energy, holographic quintom,
and the reconstruction of scalar-field dark energy. Phys Rev D,
2006, 74: 103505;
Zhang J, Zhang X, Liu H. Holographic dark energy
in a cyclic universe. Eur Phys J C, 2007, 52: 693-699;
  Setare M R, Zhang J, Zhang X.
  Statefinder diagnosis in a non-flat universe and the holographic model of dark energy.
  J Cosmol Astrpart Phys, 2007, 03: 007;
Zhang X. Heal the world: Avoiding the cosmic
doomsday in the holographic dark energy model. Phys Lett B, 2010, 683:
81-87;
  Zhang J F, Li Y Y, Liu Y, et al.
  Holographic $\Lambda$(t)CDM model in a non-flat universe.
  Eur Phys J C, 2012, 72: 2077.


\bibitem{holofit}
  Zhang X, Wu F Q. Constraints on holographic dark energy from type Ia
supernova observations. Phys Rev D, 2005, 72: 043524;
  Zhang X, Wu F Q. Constraints on holographic dark energy
from latest supernovae, galaxy clustering, and cosmic microwave background
anisotropy observations. Phys Rev D, 2007, 76: 023502;
  Huang Q G, Gong Y G. Supernova constraints on a
holographic dark energy model. J Cosmol Astropart Phys, 2004, 08:
006;
  Chang Z, Wu F Q, Zhang X. Constraints
on holographic dark energy from X-ray gas mass fraction of galaxy clusters.
Phys Lett B, 2006, 633: 14-18;
  Yi Z L, Zhang T J.
  Constraints on holographic dark energy models using the differential ages
  of passively evolving galaxies.
  Mod Phys Lett A, 2007, 22: 41-54;
Zhang X. Holographic Ricci dark energy:
Current observational constraints, quintom feature, and the reconstruction
of scalar-field dark energy. Phys Rev D, 2009, 79: 103509;
  Li M, Li X D,Wang S, et al. Holographic dark energy models: A comparison
from the latest observational data. J Cosmol Astropart Phys, 2009,
06: 036;
  Li M, Li X D,Wang S, et al.
  Probing interaction and spatial curvature in the holographic dark energy model.
  J Cosmol Astropart Phys, 2009,12: 014;
  Li X D, Wang S, Huang Q G, et al.
  Dark energy and fate of the universe.
  Sci China Phys Mech Astron, 2012, 55: 1330-1334;
  Li Y H, Wang S, Li X D, Zhang X.
  Holographic dark energy in a Universe with spatial curvature and massive neutrinos: a full Markov Chain Monte Carlo exploration.
  J Cosmol Astropart Phys, 2013, 02: 033.




\bibitem{Wei:2007nade}
  Wei H, Cai R G. A new model of agegraphic dark energy. Phys Lett B,
2008, 660: 113-117.

\bibitem{Cai:2007ade}
  Cai R G. A dark energy model characterized by the age of the universe.
Phys Lett B, 2007, 657: 228-231.



\bibitem{nadefit}
  Wei H, Cai R G. Cosmological constraints on new agegraphic dark energy.
Phys Lett B, 2008, 663: 1-6;
  Zhang J, Zhang L, Zhang X. Sandage-Loeb test for the new agegraphic
and Ricci dark energy models. Phys Lett B, 2010, 691: 11-17;
  Zhang J F, Li Y H, Zhang X.
  A global fit study on the new agegraphic dark energy model.
  Eur Phys J C, 2013, 73: 2280.


\bibitem{NADEext}
  Zhang J, Zhang X, Liu H. Agegraphic dark energy
as a quintessence. Eur Phys J C, 2008, 54: 303-309;
  Cui J, Zhang L, Zhang J, et al. New agegraphic dark
energy as a rolling tachyon. Chin Phys B, 2010, 19: 019802;
  Neupane I P. Remarks on dynamical dark energy measured by the conformal
age of the universe. Phys Rev D, 2007, 76: 123006;
  Liu X L, Zhang X. New agegraphic dark energy
in Brans-Dicke theory. Commun Theor Phys, 2009, 52: 761-768;
  Liu X L, Zhang J, Zhang X. Theoretical
limits on agegraphic quintessence from weak gravity conjecture. Phys
Lett B, 2010, 689: 139-144.



\bibitem{NADEini}
  Li Y H, Zhang J F, Zhang X.
  New initial condition of the new agegraphic dark energy model.
  Chin Phys B, 2013, 37: 039501.


\bibitem{INADEold1}
  Zhang L, Cui J, Zhang J,
et al. Interacting model of new agegraphic dark energy: Cosmological
evolution and statefinder diagnostic. Int J Mod Phys D, 2010, 19: 21-35.

\bibitem{INADEold2}
  Li Y H, Ma J Z, Cui J L, et al.
  Interacting model of new agegraphic dark energy: observational constraints and age problem.
  Sci China Phys Mech Astron, 2011, 54: 1367-1377.

\bibitem{intde}
Boehmer C G, Caldera-Cabral G, Lazkoz R, et al. 
Dynamics of dark energy with a coupling to dark matter. 
Phys Rev D, 2008, 78: 023505; 
Valiviita J, Majerotto E, Maartens R. 
Large-scale instability in interacting dark energy and dark matter fluids.
J Cosmol Astropart Phys, 2008, 07: 020; 
Caldera-Cabral G, Maartens R, Urena-Lopez L A. 
Dynamics of interacting dark energy.
Phys Rev D, 2009, 79: 063518;
Caldera-Cabral G, Maartens R, Schaefer B M.
The growth of structure in interacting dark energy models.
J Cosmol Astropart Phys, 2009, 07: 027; 
Koyama K, Maartens R, Song Y S.
Velocities as a probe of dark sector interactions.
J Cosmol Astropart Phys, 2009, 10: 017; 
Majerotto E, Valiviita J, Maartens R.
Adiabatic initial conditions for perturbations in interacting dark energy models.
Mon Not R Astron Soc, 2010, 402: 2344-2354; 
Valiviita J, Maartens R, Majerotto E.
Observational constraints on an interacting dark energy model.
Mon Not R Astron Soc, 2010, 402: 2355-2368;
Clemson T, Koyama K, Zhao G B, et al.
Interacting dark energy: Constraints and degeneracies.
Phys Rev D, 2012, 85: 043007.

\bibitem{intholo}
  Zhang Z, Li S, Li X D, et al.
  Revisit of the interaction between holographic dark energy and dark matter.
  J Cosmol Astropart Phys, 2012, 06: 009.

\bibitem{Wang:2013mha}
  Wang Y, Wang S.
  Distance priors from Planck and dark energy constraints from current data.
  Phys Rev D, 2013, 88: 043522.


\bibitem{union2.1}
  Suzuki N, Rubin D, Lidman C, et al. 
  The Hubble Space Telescope cluster supernova survey: V. Improving the dark energy constraints above $z>1$ and building an early-type-hosted supernova sample.
  Astrophys J, 2012,  746: 85.



\bibitem{Riess:2011HST}
  Riess A G, Macri L, Casertano S, et al. 
  A $3\%$ solution: Determination of the Hubble constant with the Hubble Space Telescope and Wide Field Camera 3.
  Astrophys J, 2011, 730: 119.

\bibitem{demodels}
  Li M, Li X D, Zhang X.
Comparison of dark energy models: A perspective from the latest observational
data. Sci China Phys Mech Astron, 2010, 53: 1631-1645.


\bibitem{Wei:2010BIC}
  Wei H.
  Observational constraints on cosmological models with the updated long gamma-ray bursts.
  J Cosmol Astropart Phys, 2010, 08: 020.


\bibitem{zhao2013}
Zhao L A. Some studies on the interacting model of new
agegraphic dark energy. MS Thesis (Northeastern University), 2013 (in Chinese).


\bibitem{Li:2013hdePlanck}
  Li M, Li X D, Ma Y Z, et al.
  Planck constraints on holographic dark energy.
  J Cosmol Astropart Phys, 2013, 09: 021.


\end{thebibliography}
\end{document}